\documentclass{article}
\usepackage{graphicx}
\usepackage{amsmath}
\usepackage{amsfonts}
\usepackage{amssymb}

\input epsf

\begin{document}

\title{The massless Thirring model \\in spherical field theory}
\author{Nathan Salwen\thanks{email: salwen@physics.harvard.edu}\\Harvard University\\Cambridge, MA 02138
\and Dean Lee\thanks{email: dlee@het.phast.umass.edu}\\University of Massachusetts\\Amherst, MA 01003}
\maketitle
\begin{abstract}
We use the massless Thirring model to demonstrate a new approach to
non-perturbative fermion calculations based on the spherical field formalism.
The methods we present are free from the problems of fermion doubling and
difficulties associated with integrating out massless fermions. Using a
non-perturbative regularization, we compute the two-point correlator and find
agreement with the known analytic solution.
\end{abstract}

\section{Introduction}

The massless Thirring model \cite{thirring} is an exactly soluble system of a
single self-interacting massless fermion in two dimensions. \ There are a
number of solutions\ in the literature based on properties of the
Euler-Lagrange equations and fermion currents or bosonization techniques
\cite{johnson}-\cite{mueller}. \ Given its simplicity and solubility, the
model has become a popular testing ground for new ideas and methods in field theory.

From a computational point of view, however, the massless Thirring model still
presents a significant challenge. In the lattice field formalism, the
difficulties are due to the appearance of fermion doubler states and singular
inversion problems associated with integrating out massless fermions. \ In
this work we use the model to illustrate new non-perturbative methods in the
spherical field formalism \cite{spher}-\cite{renorm}. \ The techniques we
present are quite general and can also be applied to other modal expansion
methods such as periodic field theory \cite{periodic}.

As noted in \cite{fermion}, we will not need to worry about fermion doubling.
\ This is true for any modal field theory and follows from the fact that space
is not discretized but retained as a continuous variable. \ Since our model is
not super-renormalizable we will use a procedure called angle-smearing, a
regularization method designed for spherical field theory \cite{renorm}.
\ With angle-smearing regularization and a small set of local counterterms, we
are able to remove all ultraviolet divergences in a manner such that the
renormalized theory is finite and translationally invariant. \ Comparison of
our results with the known Thirring model solution will serve as a consistency
check for our regularization and renormalization procedures.

The organization of this paper is as follows. \ We begin with a short summary
of the massless Thirring model, following the solution of Hagen \cite{hagen}.
Using angle-smearing regularization we obtain the spherical field Hamiltonian
and construct a matrix representation of the Hamiltonian. We reduce the space
of states using a two-parameter auxiliary cutoff procedure. \ In this reduced
space we compute the time evolution of quantum states using a fourth-order
Runge-Kutta-Fehlberg algorithm. \ We calculate the two point correlator for
several values of the coupling and find agreement with the known analytic solution.

\section{The model}

We start with a list of our notational conventions. \ Our analysis will be in
two-dimensional Euclidean space, and we use both cartesian and polar
coordinates,
\begin{equation}
\vec{t}=(t\cos\theta,t\sin\theta)=(x,y).
\end{equation}
The components of the spinors $\psi$ and $\bar{\psi}$ are written as
\begin{equation}
\psi=\left[
\begin{array}
[c]{c}%
\psi^{\uparrow}\\
\psi^{\downarrow}%
\end{array}
\right]  \qquad\bar{\psi}=\left[
\begin{array}
[c]{cc}%
\bar{\psi}^{\uparrow} & \bar{\psi}^{\downarrow}%
\end{array}
\right]  .
\end{equation}
Our representation for the Dirac matrices is%

\begin{equation}
\vec{\gamma}=i\vec{\sigma},
\end{equation}
and so $\vec{\gamma}$ satisfies
\begin{equation}
\left\{  \gamma^{i},\gamma^{j}\right\}  =-2\delta^{ij},\qquad i,j=1,2.
\end{equation}

The massless Thirring model is formally defined by the Lagrange density
\begin{equation}
\mathcal{L}=i\bar{\psi}\vec{\gamma}\cdot\vec{\nabla}\psi-\tfrac{\lambda}%
{2}\,\vec{j}\cdot\vec{j},
\end{equation}
where $\vec{j}$ is the fermion vector current. \ Johnson \cite{johnson}
emphasized the importance of defining the regularized current precisely, and
this was further clarified by the work of Sommerfield \cite{sommerfield} and
Hagen \cite{hagen}. We will use a regularization technique, introduced in
\cite{renorm}, called angle-smearing. We define the regularized current as
\begin{equation}
\vec{j}=\tfrac{1}{2}\left(  \bar{\psi}_{s}\vec{\gamma}\psi_{s}-Tr[\vec{\gamma
}\psi_{s}\bar{\psi}_{s}]\right)  ,
\end{equation}
where
\begin{equation}
\psi_{s}(t,\theta)=\tfrac{Mt}{2}\int_{-\frac{1}{Mt}}^{\frac{1}{Mt}%
}d\varepsilon\psi(t,\theta+\varepsilon). \label{sme}%
\end{equation}
We identify the radial variable $t$ as our time parameter, and our definition
of the current is local with respect to $t$.

Hagen \cite{hagen} described the solution of the Thirring model in the
Hamiltonian formalism with currents defined as products of the canonical
operators at equal times. Though our equal-time surface is curved, the
curvature of the integration segment in (\ref{sme}) scales as $\frac{1}{M}$
while the ultraviolet divergences in this model are only logarithmic in $M$.
In the $M\rightarrow\infty$ limit we therefore recover the standard results.
As discussed in \cite{hagen}, there exists a one parameter class of solutions
to the Thirring model depending on the preferred definition of the regularized
vector and axial vector currents. We will use the conventions used in
\cite{johnson} and \cite{sommerfield}, which in Hagen's notation corresponds
with the parameter values $\xi=\eta=\frac{1}{2}$. \ With this choice the
Hamiltonian density takes the form
\begin{equation}
\mathcal{H}=\mathcal{H}_{free}+\tfrac{\pi c}{1-c}(\hat{t}\cdot\vec{j}%
)^{2}+\tfrac{\pi c}{1+c}(\hat{\theta}\cdot\vec{j})^{2},\label{la}%
\end{equation}
where
\begin{equation}
c=\tfrac{\lambda}{2\pi}\text{.}%
\end{equation}

The only counterterms needed in this model are wavefunction renormalization
counterterms, a result of our careful definition for the regularized
interaction in (\ref{la}). As in \cite{hagen} we calculate correlation
functions using an unrenormalized Hamiltonian. \ The divergent wavefunction
normalizations will appear simply as overall factors in the correlators.\qquad

\section{Spherical field Hamiltonian}

In this section we derive the form of the spherical field Hamiltonian. We
first expand the fermion current in terms of components of the spinors,
\begin{equation}
\hat{t}\cdot\bar{\psi}_{s}\vec{\gamma}\psi_{s}=i\bar{\psi}_{s}\left[
\begin{array}
[c]{cc}%
0 & e^{-i\theta}\\
e^{i\theta} & 0
\end{array}
\right]  \psi_{s}=ie^{-i\theta}\bar{\psi}_{s}^{\uparrow}\psi_{s}^{\downarrow
}+ie^{i\theta}\bar{\psi}_{s}^{\downarrow}\psi_{s}^{\uparrow}%
\end{equation}%
\begin{equation}
\hat{\theta}\cdot\bar{\psi}_{s}\vec{\gamma}\psi_{s}=\bar{\psi}_{s}\left[
\begin{array}
[c]{cc}%
0 & e^{-i\theta}\\
-e^{i\theta} & 0
\end{array}
\right]  \psi_{s}=e^{-i\theta}\bar{\psi}_{s}^{\uparrow}\psi_{s}^{\downarrow
}-e^{i\theta}\bar{\psi}_{s}^{\downarrow}\psi_{s}^{\uparrow}.
\end{equation}
The anti-commutators of the regulated fields are\footnote{Our definition of
the Euclidean fermion fields and anti-commutation relations follows the
conventions of \cite{fubini}.}
\begin{equation}
\left\{  \bar{\psi}_{s}^{\uparrow},\psi_{s}^{\downarrow}\right\}  =\tfrac
{1}{t}\left(  \tfrac{Mt}{2}\right)  ^{2}%
{\textstyle\int_{-\frac{1}{Mt}}^{\frac{1}{Mt}}}
e^{i(\theta+\varepsilon)}d\varepsilon=A(t)e^{i\theta} \label{an1}%
\end{equation}%
\begin{equation}
\left\{  \bar{\psi}_{s}^{\downarrow},\psi_{s}^{\uparrow}\right\}  =\tfrac
{1}{t}\left(  \tfrac{Mt}{2}\right)  ^{2}%
{\textstyle\int_{-\frac{1}{Mt}}^{\frac{1}{Mt}}}
e^{-i(\theta+\varepsilon)}d\varepsilon=A(t)e^{-i\theta}, \label{an2}%
\end{equation}
where
\begin{equation}
A(t)=\tfrac{M^{2}t}{2}\sin(\tfrac{1}{Mt}). \label{a}%
\end{equation}
From the anti-commutation relations, the $\hat{t}$ component of the current
is
\begin{align}
\hat{t}\cdot\vec{j}  &  =\tfrac{1}{2}\left[  ie^{-i\theta}(\bar{\psi}%
_{s}^{\uparrow}\psi_{s}^{\downarrow}-\psi_{s}^{\downarrow}\bar{\psi}%
_{s}^{\uparrow})+ie^{i\theta}(\bar{\psi}_{s}^{\downarrow}\psi_{s}^{\uparrow
}-\psi_{s}^{\uparrow}\bar{\psi}_{s}^{\downarrow})\right] \\
&  =ie^{-i\theta}\bar{\psi}_{s}^{\uparrow}\psi_{s}^{\downarrow}+ie^{i\theta
}\bar{\psi}_{s}^{\downarrow}\psi_{s}^{\uparrow}-iA(t),\nonumber
\end{align}
and so
\begin{equation}
(\hat{t}\cdot\vec{j})^{2}=A(t)\left[  e^{-i\theta}\bar{\psi}_{s}^{\uparrow
}\psi_{s}^{\downarrow}+e^{i\theta}\bar{\psi}_{s}^{\downarrow}\psi
_{s}^{\uparrow}\right]  -2\bar{\psi}_{s}^{\uparrow}\psi_{s}^{\downarrow}%
\bar{\psi}_{s}^{\downarrow}\psi_{s}^{\uparrow}-A^{2}(t).
\end{equation}
Similarly we find
\begin{equation}
(\hat{\theta}\cdot\vec{j})^{2}=A(t)\left[  e^{-i\theta}\bar{\psi}%
_{s}^{\uparrow}\psi_{s}^{\downarrow}+e^{i\theta}\bar{\psi}_{s}^{\downarrow
}\psi_{s}^{\uparrow}\right]  -2\bar{\psi}_{s}^{\uparrow}\psi_{s}^{\downarrow
}\bar{\psi}_{s}^{\downarrow}\psi_{s}^{\uparrow}.
\end{equation}
The Hamiltonian can now be written as
\begin{equation}
H=H_{free}+\int d\theta\,t\left\{  \tfrac{2\pi c}{1-c^{2}}\left[  A(t)\left[
e^{-i\theta}\bar{\psi}_{s}^{\uparrow}\psi_{s}^{\downarrow}+e^{i\theta}%
\bar{\psi}_{s}^{\downarrow}\psi_{s}^{\uparrow}\right]  -2\bar{\psi}%
_{s}^{\uparrow}\psi_{s}^{\downarrow}\bar{\psi}_{s}^{\downarrow}\psi
_{s}^{\uparrow}\right]  \right\}  .
\end{equation}
We have omitted the constant term proportional to $A^{2}(t)$.

Let us define the partial wave modes
\begin{align}
\psi_{n}(t)  &  =\tfrac{1}{\sqrt{2\pi}}\int d\theta\,e^{-in\theta}\psi(\vec
{t}),\qquad\psi_{s,n}(t)=\tfrac{1}{\sqrt{2\pi}}\int d\theta\,e^{-in\theta}%
\psi_{s}(\vec{t}),\\
\bar{\psi}_{n}(t)  &  =\tfrac{1}{\sqrt{2\pi}}\int d\theta\,e^{-in\theta}%
\bar{\psi}(\vec{t}),\qquad\bar{\psi}_{s,n}(t)=\tfrac{1}{\sqrt{2\pi}}\int
d\theta\,e^{-in\theta}\bar{\psi}_{s}(\vec{t}).
\end{align}
It is straightforward to show that for $n\neq0,$
\begin{equation}
\psi_{s,n}(t)=\tfrac{\sin(\frac{n}{Mt})}{\left(  \frac{n}{Mt}\right)  }%
\psi_{n}(t)\qquad\bar{\psi}_{s,n}(t)=\tfrac{\sin(\frac{n}{Mt})}{\left(
\frac{n}{Mt}\right)  }\bar{\psi}_{n}(t). \label{a1}%
\end{equation}
We can extend this result to the case $n=0$ using the convenient shorthand
\begin{equation}
\tfrac{\sin(\frac{0}{Mt})}{\left(  \frac{0}{Mt}\right)  }\equiv1.
\end{equation}
At this point it is convenient to rescale $\bar{\psi}$,
\begin{equation}
\bar{\psi}_{n}^{i\prime}=t\bar{\psi}_{n}^{i}.
\end{equation}
In terms of the partial waves,
\begin{align}
H=  &  \tfrac{1}{t}\sum_{n}\left[  \left(  n+1+\tfrac{b\pi tA(t)\sin(\frac
{n}{Mt})\sin(\frac{n+1}{Mt})}{\left(  \frac{n}{Mt}\right)  \left(  \frac
{n+1}{Mt}\right)  }\right)  \bar{\psi}_{-n}^{\uparrow\prime}\psi
_{n+1}^{\downarrow}\right] \\
&  -\tfrac{1}{t}\sum_{n}\left[  \left(  n-\tfrac{b\pi tA(t)\sin(\frac{n}%
{Mt})\sin(\frac{n+1}{Mt})}{\left(  \frac{n}{Mt}\right)  \left(  \frac{n+1}%
{Mt}\right)  }\right)  \bar{\psi}_{-n-1}^{\downarrow\prime}\psi_{n}^{\uparrow
}\right] \nonumber\\
&  -\sum_{-n_{1}+n_{2}-n_{3}+n_{4}=0}\left[  \tfrac{b}{t}\bar{\psi}_{-n_{1}%
}^{\uparrow\prime}\psi_{n_{2}+1}^{\downarrow}\bar{\psi}_{-n_{3}-1}%
^{\downarrow\prime}\psi_{n_{4}}^{\uparrow}\tfrac{\sin(\frac{n_{1}}{Mt}%
)\sin(\frac{n_{2}+1}{Mt})\sin(\frac{n_{3}+1}{Mt})\sin(\frac{n_{4}}{Mt}%
)}{\left(  \frac{n_{1}}{Mt}\right)  \left(  \frac{n_{2}+1}{Mt}\right)  \left(
\frac{n_{3}+1}{Mt}\right)  \left(  \frac{n_{4}}{Mt}\right)  }\right]
,\nonumber
\end{align}
where
\begin{equation}
b=\tfrac{2c}{1-c^{2}}. \label{bc}%
\end{equation}
Since $b$ is the parameter appearing in the Hamiltonian, it is somewhat more
convenient to express $c$ in terms of $b,$%
\begin{equation}
c=\tfrac{\sqrt{1+b^{2}}-1}{b}. \label{cb}%
\end{equation}

Let us define the ladder operators\footnote{This notation is slightly
different from that used in \cite{fermion}. The translation is as follows:
$a_{n}^{\downarrow},a_{n}^{\downarrow\dagger}=a_{n}^{\downarrow-}%
,a_{n}^{\downarrow+}$; $a_{n}^{\uparrow},a_{n}^{\uparrow\dagger}%
=a_{-n}^{\uparrow-},a_{-n}^{\uparrow+}.$}
\begin{align}
a_{-n}^{\uparrow},a_{-n}^{\uparrow\dagger}  &  \equiv\psi_{n}^{\uparrow}%
,\bar{\psi}_{-n-1}^{\downarrow\prime}\\
a_{n+1}^{\downarrow},a_{n+1}^{\downarrow\dagger}  &  \equiv\psi_{n+1}%
^{\downarrow},\bar{\psi}_{-n}^{\uparrow\prime}.
\end{align}
These operators satisfy the anti-commutation relations%

\begin{equation}
\left\{  a_{n_{1}}^{\downarrow},a_{n_{2}}^{\downarrow\dagger}\right\}
=\left\{  a_{n_{1}}^{\uparrow},a_{n_{2}}^{\uparrow\dagger}\right\}
=\delta_{n_{1}n_{2}},
\end{equation}
with all other anti-commutators equal to zero. We can now recast the
Hamiltonian as
\begin{align}
H  &  =\tfrac{1}{t}\sum_{n}\left[  n+\tfrac{b\pi tA(t)\sin(\frac{n}{Mt}%
)\sin(\frac{n-1}{Mt})}{\left(  \frac{n}{Mt}\right)  \left(  \frac{n-1}%
{Mt}\right)  }\right]  \left(  a_{n}^{\downarrow\dagger}a_{n}^{\downarrow
}+a_{n}^{\uparrow\dagger}a_{n}^{\uparrow}\right) \label{ha}\\
&  -\sum_{-n_{1}+n_{2}+n_{3}-n_{4}=0}\left[  \tfrac{b}{t}a_{n_{1}}%
^{\downarrow\dagger}a_{n_{2}}^{\downarrow}a_{n_{3}}^{\uparrow\dagger}a_{n_{4}%
}^{\uparrow}\tfrac{\sin(\frac{n_{1}-1}{Mt})\sin(\frac{n_{2}}{Mt})\sin
(\frac{n_{3}-1}{Mt})\sin(\frac{n_{4}}{Mt})}{\left(  \frac{n_{1}-1}{Mt}\right)
\left(  \frac{n_{2}}{Mt}\right)  \left(  \frac{n_{3}-1}{Mt}\right)  \left(
\frac{n_{4}}{Mt}\right)  }\right]  .\nonumber
\end{align}

We will implement a high spin cutoff by removing terms in the interaction
containing operators $a_{n}^{\downarrow}$, $a_{n}^{\uparrow},$ $a_{n}%
^{\downarrow\dagger}$, or $a_{n}^{\uparrow\dagger}$ for $\left|  n\right|
>J_{\max}$. This has the effect of removing high spin modes, which correspond
with large tangential momentum states. \ We should emphasize, however, that
$J_{\max}$ is an auxiliary cutoff. It does not play a role in the
regularization scheme since the interactions have already been rendered finite
using angle-smearing. \ The contribution of these high spin modes is
negligible so long as
\begin{equation}
\tfrac{J_{\max}}{t}\gg M,
\end{equation}
where $t$ is the characteristic radius of the process being measured.
\ Returning back to (\ref{an1}) and (\ref{an2}) and removing the contribution
of these partial waves, we find that $A(t)$ is replaced by
\begin{equation}
A_{J_{\max}}(t)=\tfrac{1}{2\pi t}\sum_{n=-J_{\max}}^{J_{\max}}\tfrac
{\sin(\frac{n}{Mt})}{\left(  \frac{n}{Mt}\right)  }\tfrac{\sin(\frac{n-1}%
{Mt})}{\left(  \frac{n-1}{Mt}\right)  }. \label{sum}%
\end{equation}

Let $\left|  0\right\rangle _{free}$ be the ground state of the free massless
fermion Hamiltonian.\footnote{The ground state of the free massless
Hamiltonian is actually degenerate due to s-wave excitations, but this is
remedied by taking the $m\longrightarrow0$ limit of the massive fermion
theory.} For $n>0,$ we find%

\begin{equation}
a_{n}^{\downarrow}\left|  0\right\rangle _{free}=a_{n}^{\uparrow}\left|
0\right\rangle _{free}=0,
\end{equation}
and for $n\leq0,$
\begin{equation}
a_{n}^{\downarrow\dagger}\left|  0\right\rangle _{free}=a_{n}^{\uparrow
\dagger}\left|  0\right\rangle _{free}=0.
\end{equation}
It is convenient to define the normal-ordered products
\begin{equation}
\text{{}}\text{:{}}a_{n}^{\downarrow\dagger}a_{n}^{\downarrow}\text{:}%
=\left\{
\begin{array}
[c]{c}%
a_{n}^{\downarrow\dagger}a_{n}^{\downarrow}\text{ for }n>0\\
-a_{n}^{\downarrow}a_{n}^{\downarrow\dagger}\text{ for }n\leq0
\end{array}
\right.  \qquad\text{:{}}a_{n}^{\uparrow\dagger}a_{n}^{\uparrow}%
\text{:}=\left\{
\begin{array}
[c]{c}%
a_{n}^{\uparrow\dagger}a_{n}^{\uparrow}\text{ for }n>0\\
-a_{n}^{\uparrow}a_{n}^{\uparrow\dagger}\text{ for }n\leq0.
\end{array}
\right.
\end{equation}
The ordering for other operators is immaterial since the anti-commutators are
zero. We can now rewrite $H$ in terms of normal-ordered products,%

\begin{align}
H  &  =\left(  \tfrac{n}{t}+O(J_{\max}^{-2})\right)  \left(  a_{n}%
^{\downarrow\dagger}a_{n}^{\downarrow}+a_{n}^{\uparrow\dagger}a_{n}^{\uparrow
}\right) \\
&  -\sum_{-n_{1}+n_{2}+n_{3}-n_{4}=0}\left[  \tfrac{b}{t}\text{:}a_{n_{1}%
}^{\downarrow\dagger}a_{n_{2}}^{\downarrow}a_{n_{3}}^{\uparrow\dagger}%
a_{n_{4}}^{\uparrow}\text{:}\tfrac{\sin(\frac{n_{1}-1}{Mt})\sin(\frac{n_{2}%
}{Mt})\sin(\frac{n_{3}-1}{Mt})\sin(\frac{n_{4}}{Mt})}{\left(  \frac{n_{1}%
-1}{Mt}\right)  \left(  \frac{n_{2}}{Mt}\right)  \left(  \frac{n_{3}-1}%
{Mt}\right)  \left(  \frac{n_{4}}{Mt}\right)  }\right]  .\nonumber
\end{align}
There is an $O(J_{\max}^{-2})$ term due to a small asymmetry in our cutoff
procedure with respect to the two boundaries $-J_{\max}$ and $J_{\max}%
$.\footnote{If desired we could eliminate this term and the asymmetry by a
slight change in the angle-smearing procedure for $\bar{\psi}$.} We will
neglect this term in the limit $J_{\max}\rightarrow\infty$.

\section{Two-point correlator}

We wish to study the properties of the two-point correlator. The massless
Thirring model is invariant under the discrete transformation%

\begin{equation}
\psi^{\downarrow}(\vec{t}),\bar{\psi}^{\uparrow}(\vec{t})\rightarrow
-\psi^{\downarrow}(\vec{t}),-\bar{\psi}^{\uparrow}(\vec{t}),
\end{equation}
as well as the transformation
\begin{equation}
\psi^{\uparrow}(t,\theta),\bar{\psi}^{\uparrow}(t,\theta)\leftrightarrow
\psi^{\downarrow}(t,-\theta),\bar{\psi}^{\downarrow}(t,-\theta).
\end{equation}
From these we deduce
\begin{equation}
\left\langle 0\right|  T\left[  \bar{\psi}^{\uparrow}(\vec{t})\psi^{\uparrow
}(0)\right]  \left|  0\right\rangle =\left\langle 0\right|  T\left[  \bar
{\psi}^{\downarrow}(\vec{t})\psi^{\downarrow}(0)\right]  \left|
0\right\rangle =0
\end{equation}
and
\begin{equation}
\left\langle 0\right|  T\left[  \bar{\psi}^{\uparrow}(t,\theta)\psi
^{\downarrow}(0)\right]  \left|  0\right\rangle =\left\langle 0\right|
T\left[  \bar{\psi}^{\downarrow}(t,-\theta)\psi^{\uparrow}(0)\right]  \left|
0\right\rangle . \label{nn}%
\end{equation}
It therefore suffices to consider just the correlator on the left side of
(\ref{nn}).

In the limit $M\rightarrow\infty$ the form of the correlator is given by
\begin{equation}
\left\langle 0\right|  T\left[  \bar{\psi}^{\uparrow}(t,\theta)\psi
^{\downarrow}(0)\right]  \left|  0\right\rangle =\tfrac{e^{i\theta}}{2\pi
}(k(c)M)^{\frac{-2c^{2}}{1-c^{2}}}t^{\frac{-1-c^{2}}{1-c^{2}}},\label{exact}%
\end{equation}
where $k(c)$ is a dimensionless parameter. \ Standard analytic methods do not
yield a simple closed form expression for $k(c).$ \ We will therefore extract
$k(c)$ from the computed value of the correlator at a specific renormalization
point $t=t_{0}$.\footnote{In some regularization schemes $k(c)$ can be
calculated analytically \cite{sommerfield}\cite{mueller}, and it may be
worthwhile to use these techniques in future work. \ In this first analysis,
however, we prefer to present a more straightforward and typical example of
the angle-smearing regularization method.}

We define
\begin{equation}
f(t)=\left\langle 0\right|  T\left[  \bar{\psi}_{1}^{\uparrow}(t)\psi
_{0}^{\downarrow}(0)\right]  \left|  0\right\rangle .
\end{equation}
Since%

\begin{equation}
\left\langle 0\right|  T\left[  \bar{\psi}^{\uparrow}(t,\theta)\psi
^{\downarrow}(0)\right]  \left|  0\right\rangle =\tfrac{e^{i\theta}}{2\pi
}\left\langle 0\right|  T\left[  \bar{\psi}_{1}^{\uparrow}(t)\psi
_{0}^{\downarrow}(0)\right]  \left|  0\right\rangle ,
\end{equation}
we conclude that
\begin{equation}
f(t)=(k(c)M)^{\frac{-2c^{2}}{1-c^{2}}}t^{\frac{-1-c^{2}}{1-c^{2}}}.
\label{analyt}%
\end{equation}
\qquad

We now compute $f(t)$ using the spherical field Hamiltonian. \ We first need a
matrix representation for the Grassmann ladder operators. \ We will use tensor
products of the $2\times2$ identity matrix and Pauli matrices: %

\begin{align}
a_{n}^{\downarrow} &  =%
{\textstyle\bigotimes\limits_{i=J_{\max},-J_{\max}}}
\sigma_{z}%
{\textstyle\bigotimes\limits_{i=J_{\max},n+1}}
\sigma_{z}\otimes\left(  \tfrac{1}{2}\sigma_{x}+\tfrac{i}{2}\sigma
_{y}\right)
{\textstyle\bigotimes\limits_{i=n-1,-J_{\max}}}
1,\label{rep}\\
a_{n}^{\uparrow} &  =%
{\textstyle\bigotimes\limits_{i=J_{\max},n+1}}
\sigma_{z}\otimes\left(  \tfrac{1}{2}\sigma_{x}+\tfrac{i}{2}\sigma
_{y}\right)
{\textstyle\bigotimes\limits_{i=n-1,-J_{\max}}}
1%
{\textstyle\bigotimes\limits_{i=J_{\max},-J_{\max}}}
1.\nonumber
\end{align}
The representations for $a_{n}^{\downarrow\dagger}$ and $a_{n}^{\uparrow
\dagger}$ are defined by the conjugate transposes of these matrices. \ We can
now calculate the correlator $f(t)$ using the relation \cite{fermion}
\begin{equation}
f(t)=\lim_{t_{2}\rightarrow\infty}\lim_{t_{1}\rightarrow0}\tfrac{Tr\left[
T\exp\left\{  -\int_{t}^{t_{2}}dt\,H(t)\right\}  \frac{1}{t}a_{0}%
^{\downarrow\dagger}T\exp\left\{  -\int_{t_{1}}^{t}dt\,H(t)\right\}
a_{0}^{\downarrow}\right]  }{Tr\left[  T\exp\left\{  -\int_{t_{1}}^{t_{2}%
}dt\,H(t)\right\}  \right]  }.\label{ex}%
\end{equation}
A straightforward calculation of (\ref{ex}), however, is rather inefficient.
There are several techniques which we will first use to simplify the calculation.

The time evolution of the system at large $t$ is dominated by the contribution
of the ground state or, more precisely, the adiabatic flow of the
$t$-dependent ground state. \ As discussed in \cite{spher} a similar
phenomenon occurs at small $t$, due to the divergence of energy levels near
$t=0.$ \ It is therefore not necessary to compute the full matrix trace in the
numerator and denominator of (\ref{ex}). \ It is instead sufficient to compute
the corresponding ratio for a single matrix element. \ After making this
reduction, we can then go a step further and eliminate states which do not
contribute to the matrix element.

The high spin parameter $J_{\max}$ was used to remove high spin modes with
$\left|  n\right|  >J_{\max}$. \ This, however, is not a uniform cutoff in the
space of states and most of the remaining states are still high kinetic energy
states. \ Although none of the individual modes are energetic, many of the
modes can be simultaneously excited. \ Let us define $N_{n}^{\downarrow}$ and
$N_{n}^{\uparrow}$ to be bit switches, 1 or 0, depending on whether or not the
corresponding mode is excited. \ Let us also define a cutoff parameter
$K_{\max}$. We will remove all high kinetic energy states such that
\begin{equation}%
{\textstyle\sum_{n}}
\left[  \left|  n\right|  (N_{n}^{\downarrow}+N_{n}^{\uparrow})\right]  \geq
K_{\max}.
\end{equation}
For consistency $K_{\max}$ should be about the same size as $J_{\max}$.

\section{Results}

The CPU time and memory requirement for calculating (\ref{ex}) scales linearly
with the number of transitions in $H$ (i.e., non-zero elements in our matrix
representation). \ In Table 1 we have shown the number of states and
transitions for different values of $J_{\max}$.%

\[
\overset{\text{Table 1}}{%
\begin{tabular}
[c]{|l|l|l|l|l|l|l|}\hline
$J_{\max}$ & 2 & 4 & 6 & 8 & 10 & 12\\\hline
$\text{states}$ & 6 & 40 & 210 & 920 & 3600 & 13000\\\hline
transitions & 38 & 500 & 4200 & 26000 & 1.4E5 & 3.9E5\\\hline
\end{tabular}
}%
\]
We have calculated $f(t)$ for $J_{\max}\leq12$ and several values of the
coupling $b$. \ The total run time was about 100 hours on a 350 MHz PC with
256 MB RAM.

The matrix time evolution equations in (\ref{ex}) were computed using a
fourth-order Runge-Kutta-Fehlberg algorithm. \ We have set
\begin{equation}
K_{\max}=2J_{\max}+2.
\end{equation}
We will use the notation $f_{J_{\max}}(t)$ to identify the corresponding
result for a given value of $J_{\max}$. \ In Figure 1 we have plotted
$f_{J_{\max}}(t)$ for $b=1$ and $J_{\max}=4,8,12$. \ We have scaled $t$ and
$f(t)$ in dimensional units chosen such that $M=3$. \ For finite $J_{\max}$ we
expect deviations from the $J_{\max}\rightarrow\infty$ limit to be of size
$O$($\frac{M^{2}t^{2}}{J_{\max}^{2}})$. \ The curves shown in Figure 1 appear
consistent with this rate of convergence.

We can extrapolate to the limit $J_{\max}\rightarrow\infty$ using the
asymptotic form
\begin{equation}
f_{J_{\max}}(t)=f_{\infty}(t)+J_{\max}^{-2}f^{(2)}(t)+\cdots.
\end{equation}
For $b=0,0.5,1.0,3.0$ and $M=3$ we have calculated $f(t)$ using this
extrapolation technique for $J_{\max}=10$ and 12.\footnote{Both our results
and the analytic solution are even in $b$, and so we consider only positive
values.} The results are shown in Figure 2. For comparison we have plotted the
analytic solution
\begin{equation}
f_{A}(t)=(k(c)M)^{\frac{-2c^{2}}{1-c^{2}}}t^{\frac{-1-c^{2}}{1-c^{2}}%
}.\label{analyt2}%
\end{equation}
The relation between $b$ and $c$ can be found in (\ref{bc}) and (\ref{cb}).
\ The parameter $k(c)$ is fitted to the value of the correlator at the
renormalization point $t=0.6.\footnote{The relative error is expected to be
small in the vicinity of this point.}$ The agreement appears quite good.
\ Some deviations from the analytic solution are due to $O(\frac{1}{M^{2}%
t^{2}})$ residual terms, which were left out of the derivation of
(\ref{analyt2}). \ These effects are significant in the small $t$ region,
$t\lesssim M^{-1}$, especially for larger values of $b$. \ The values we find
for $k(c)$ are shown in Table 2.$\footnote{In some regularization schemes
$k(c)$ can be shown to be independent of the coupling. \ Our regularization
method seems to be rather close to this, with only a slow variation with
respect to the coupling strength.}$%
\[
\overset{\text{Table 2}}{%
\begin{tabular}
[c]{|l|l|l|l|}\hline
$b$ & 0.5 & 1.0 & 3.0\\\hline
$k(c)$ & 1.77 & 1.77 & 1.68\\\hline
\end{tabular}
}%
\]
We can compare our results at small $b$ with a simple perturbative
calculation. Evaluating the corresponding regulated two-loop diagram we
obtain, for small $b$, $k(c)\approx1.75$.  This appears consistent with the
results in Table 2.

\section{Summary}

We derived the angle-smeared spherical Hamiltonian for the massless Thirring
model and constructed an explicit matrix representation. We discarded
negligible high energy states using auxiliary cutoff parameters $J_{\max}$ and
$K_{\max}$. \ In this reduced space we computed the time evolution of quantum
states and calculated the two-point correlator for several values of the
coupling. The results of our computation are in close agreement with the known
analytic solution. \ In addition to demonstrating new computational methods,
our analysis also serves as a consistency check of the regularization and
renormalization methods introduced in \cite{renorm}.

We believe that this work represents a significant new direction in the
non-perturbative computation of fermion dynamics. \ Future work will study the
application of these methods to systems of interacting bosons and fermions. \ 

{\normalsize \bigskip}

\noindent{\Large \textbf{Acknowledgment}}{\normalsize \bigskip}

\noindent We thank C. R. Hagen and R. Jackiw for useful comments and
correspondence regarding the massless Thirring model and acknowledge financial
support provided by the NSF under Grant 5-22968 and PHY-9802709.

\noindent{\Large \textbf{Figures}}\bigskip

\noindent Figure 1. Plot of $f_{J_{\max}}(t)$ for $b=1$ and $J_{\max}=4,8,12$.

\noindent Figure 2. Plot of $f_{A}(t)$ and $f(t)$ for $b=0,.5,1,3$ and $M=3$.

\newpage

\begin{figure}[ptb]
\epsfbox{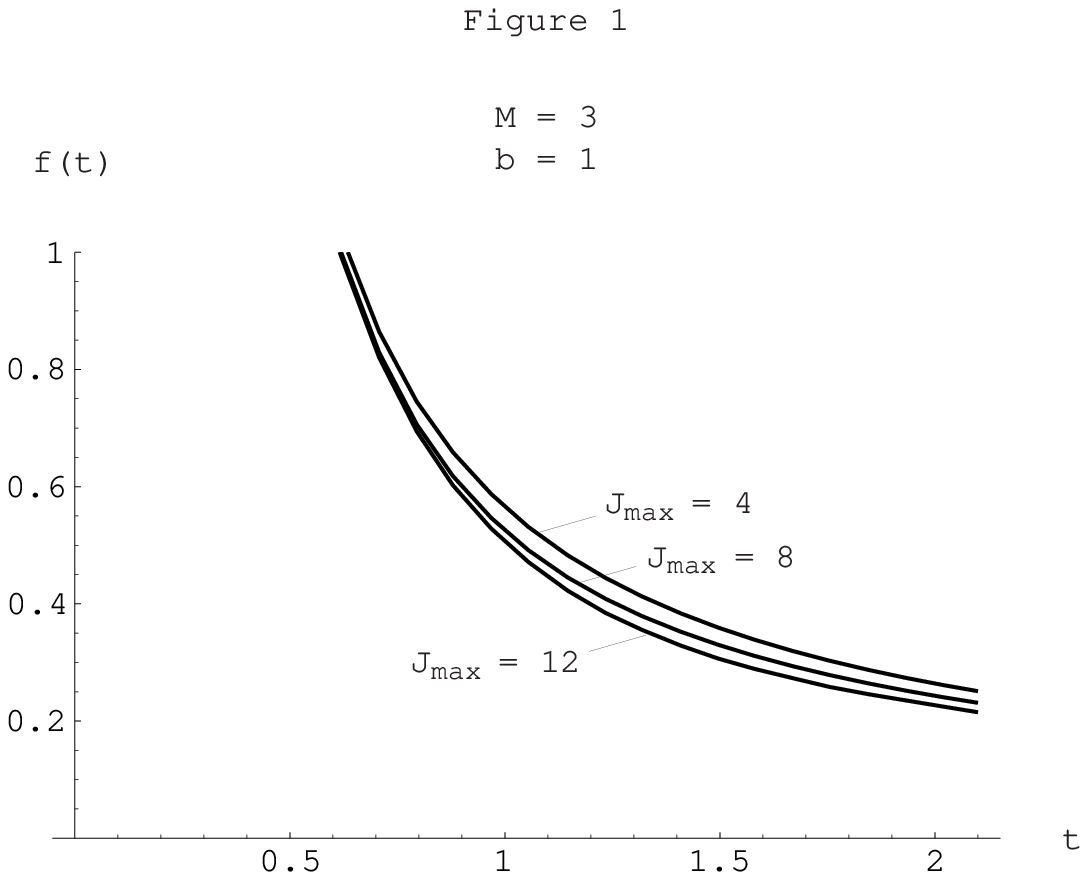}\end{figure}\begin{figure}[ptbptb]
\epsfbox{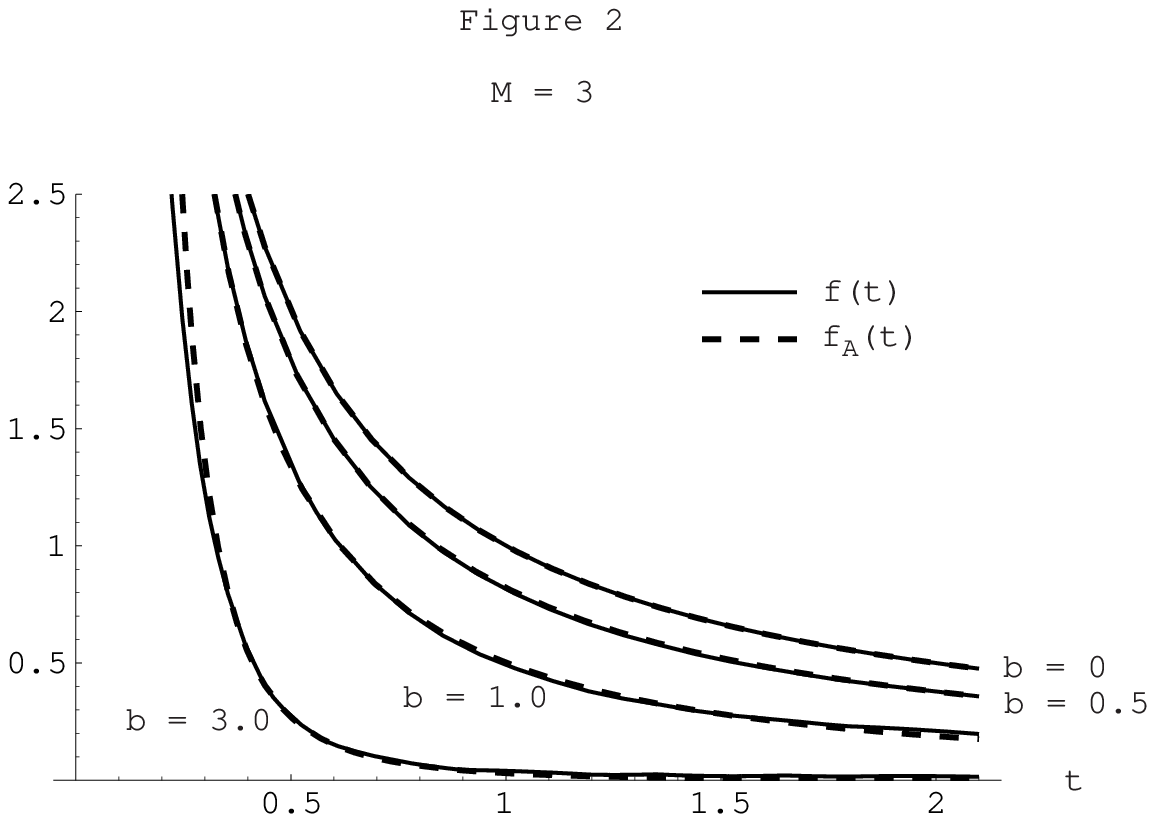}\end{figure}
\end{document}